# Adsorption of a Cationic Laser Dye onto Polymer/Surfactant Complex Film Fabricated by Layer-by-Layer Electrostatic Self-Assembly Technique

P. K. Paul[a*], S. A. Hussain[b], D. Bhattacharjee[b] and M. Pal[c]

a. Department of Physics, Jadavpur University, Jadavpur, Kolkata-700032, West Bengal, India
b. Department of Physics, Tripura University, Suryamaninagar-799130, Tripura West, India
c. Centre for Advanced Materials Processing (CAMP), Central Mechanical Engineering Research Institute (CMERI), Mahatma Gandhi Avenue, Durgapur - 713209, India.

**Abstract:** Fabrication of complex molecular films of organic materials is one of the most important issues in modern nanoscience and nanotechnology. Soft materials with flexible properties have been given much attention and can be obtained through bottom up processing from functional molecules, where self-assembly based on supramolecular chemistry and designed assembly have become crucial processes and technologies. In this short communication, we report the successful incorporation of cationic laser dye Rhodamine 6G abbreviated as R6G into the pre-assembled polyelectrolyte/Surfactant complex film onto quartz substrate by electrostatic adsorption technique. Poly(allylamine hydrochloride) (PAH) was used as polycation and Sodium dodecyl sulphate(SDS) was used as anionic surfactant. UV-Vis absorption spectroscopic characterization reveals the formation of only H-type aggregates of R6G in their aqueous solution and both H and J-type aggregates in PAH/SDS/R6G complex LbL films as well as the kinetics of adsorption of R6G onto the complex films. The ratio of the absorbance intensity of two aggregated bands in PAH/SDS/R6G complex films is merely independent of the concentration range of the SDS solution used to fabricate PAH/SDS complex self-assembled films. Atomic Force Microscopy reveals the formation of R6G aggregates in PAH/SDS/R6G complex films.

## 1. Introduction

Layer-by-Layer Electrostatic self-assembly technique is a versatile approach for the fabrication of nanoscale thin films of organic dyes onto solid substrates. However, the aggregation and other physicochemical properties of the adsorbed dye can be controlled with ease by the incorporation of some oppositely charged surfactant in the assembly. In recent times polyelectrolyte-surfactant complexation and their interactions with some cationic laser dye at the solid-liquid interface has drawn much attention in the field of material chemistry [1]. Such complexes posses growing commercial relevance as materials for separation-membranes, solubilization and compatibilization [2].

In the present work we have addressed the formation of polymer-surfactant complex architectures onto quartz substrate and thereby successful incorporation of a cationic laser dye onto this complex assemblies to fabricate polymer/surfactant/dye complex molecular films by electrostatic layer-by-layer self-assembly technique. Poly(allylamine hydrochloride) (PAH) was taken as the cationic polyelectrolyte and Sodium dodecyl sulphate (SDS) was taken as anionic surfactant. The dye molecule studied here was Rhodamine 6G (R6G) and it is a cationic laser dye. Rhodamine 6G (R6G) is a dye from the xanthene family having a molecular structure as shown in figure 1(a). R6G have been used extensively as a sensor [3], nonlinear optical material [4] and photosensitizer [5]. R6G had also been utilized as a probe molecule in the field of extremely sensitive detection, such as single-molecule detection using the surface Enhanced Resonance Raman Effect [6,7], nonlinear vibrational detection using hyper-Raman scattering [8] and nanometer scale detection using near-field Raman spectroscopy [9].

Therefore, it is extremely important to understand the photophysical and photochemical behaviour of the adsorbed dye to the polymer-surfactant complex architectures onto solid substrates. It is worthy to mention that dye-surfactant interactions are of great interest in dyeing and photographic industries [10] in biological and medicinal photosensitization [11].

Owing to the displacement of small counterions, polyelectrolyte-surfactant association can be both entropically and electrostatically driven with a modest contribution from hydrophobic interactions [12]. The expulsion of small counterions into the solvent during ion-pair complexation is the key driving force for many ionic aggregation processes involving macromolecules. The ionic character of these aggregates favours the adsorption of oppositely charged dye molecules to form polymer/surfactant/dye complex molecular systems onto the solid support.

The purpose of this work is to understand the photophysical behaviour of the PAH/SDS/R6G complex LbL films as well as also the kinetics of adsorption of R6G to the PAH/SDS complex architectures in a narrow concentration range of SDS where the system may be sensitive to surfactant concentration. This system was selected among several others (eosin or RB with CTAB and MB with SDS) because of the interesting photophysical and photochemical properties of the monomeric and the aggregated dye [13]. Both H- and J-type aggregates of R6G are formed in PAH/SDS/R6G complex LbL films as evidenced from their spectroscopic characterization.

## 2. Experimental
## 2.1 Materials

The Cationic laser dye used in the present work was Rhodamine 6G(R6G) ($C_{28}H_{31}N_2O_3Cl$, MW= 479.0). Cationic polymer poly(allylamine hydrochloride) (PAH) (molecular weight = 70,000) and anionic surfactant Sodium dodecyl Sulphate (SDS) ($CH_3(CH_2)_{11}OSO_3Na$, MW= 288.38) were also used in this work. SDS (Purity >99%) was purchased from BDH Chemical Co, England. R6G dye and the polymer PAH (Purity >99%) were purchased from Aldrich Chemical Co., USA and were used without any further purification. Chemical structures of these compounds are shown in figure 1(a), 1(b) and 1(c). The film depositions were done onto the thoroughly cleaned fluorescence grade quartz substrates. The electrolytic deposition baths were prepared with $10^{-3}$ M (based on repeat units for the polyion) aqueous solutions. Pure water was taken from a Millipore system comprising reverse osmosis followed by ion exchange and filtration steps.

## 2.2 Preparation of PAH/SDS/R6G complex LbL films

The detail experimental procedure for fabricating LbL electrostatic self-assembled films has been described by several authors [14, 15]. In this experiment PAH was deposited onto quartz substrate for 15 min. Then after rinsing with pure water, the film containing PAH layer was immersed into anionic SDS solutions having different varying concentrations ($10^{-2}$ M -$10^{-5}$ M) for 15 min. PAH was used here to fix the anionic surfactant SDS molecules onto the top of the PAH layer so that the terminal surface becomes negatively charged. Then each of these LbL self-assembled films of PAH/SDS systems onto quartz substrates were immersed into the R6G aqueous solutions having dye concentration $10^{-5}$M. Sufficient time was allowed to adsorb the dye molecules onto PAH/SDS LbL films and thus the PAH/SDS/R6G complex LbL self assembled films were prepared. The adsorption of the R6G molecules on the PAH/SDS systems as well as the morphology of the complex film were studied by of UV-Vis absorption spectroscopy and Atomic Force Microscopy [Veeco, Digital Instrument CP II Microscope] respectively.

## 3. Results and discussion

Figure 2 shows the normalized UV-Vis absorption spectra of R6G aqueous solution. From the figure it is observed that the electronic absorption spectra of R6G aqueous solutions have the maximum absorption band at around 526 nm which is due to R6G monomer and a higher energy shoulder band at around 495 nm due to the aggregates of R6G (sometimes called H dimer) [16]. These dimers are formed through van der Waals dye-dye interactions and Rhodamine-water (counterions) interactions [17]. It may be mentioned in this context that in order to check the contribution of H–dimer we measured the absorption spectra at high dye concentration in solution. Little increase of relative intensity of 495 nm band in comparison to 526 nm band was observed (figure not shown). This may be due to the contribution of aggregate and consequent formation of H–dimer. It is also relevant to mention in this context that in several other works using R6G dye high energy shoulder of H–dimer along with monomeric band was observed [18].

The molar extinction coefficient ($\varepsilon$) was calculated as 81,000 mol$^{-1}$cm$^{-1}$ at wavelength 526 nm. The presence of two bands in the absorption spectra of the dye solution agreed completely with existing concept of the splitting of energy levels upon combination of the dye molecules into H-type species such as dimer, trimer or higher order aggregates with the increase in molar concentration. Baranova and Levshin reported that Rhodamine 6G aggregates in the dimerization stage at concentration bellow $2\times10^{-3}$ mole/l but higher aggregates are formed at higher concentration [19]. At the low concentration range the formation of the dimer and their hypsochromic shift when compared to their monomeric band has already been interpreted in terms of weak coupling theory as suggested by De Voe [20].

The experimental data suggests that the absorption of R6G associates has a minimum in the range 525-544 nm; therefore the contribution of absorption of molecular aggregates of this compound to the overall electronic absorption spectrum of the solution was small in this spectral range [21] for very low dye concentration. This is why the LbL film fabrication was performed using very low concentration i.e. $10^{-5}$ M of R6G in water.

In the present work we have basically examined the successful incorporation of the cationic laser dye R6G onto the pre-assembled PAH/SDS molecular films. Figure 3 shows the absorption spectra of PAH/SDS/R6G complex Layer-by-Layer self-assembled films onto quartz substrate for different varying concentrations ($10^{-2}$ M - $10^{-5}$ M) of SDS aqueous solutions. In all the cases 15 min were considered for the deposition of SDS and R6G in the complex LbL films. From the figure it is observed that the higher and lower energy bands have been shifted to 509 nm 540 nm respectively when compared to their solution absorption spectrum. Also the spectrum is over all broadened with respect to their solution counter part. The occurrence of 509 nm and 540 nm band in complex LbL films and the resultant red shift of the absorption spectra of LbL films when compared to absorption spectrum of R6G solution are due to the formation of greater number of H-type dimer aggregates along with formation of J-type aggregates. It is these J-type aggregates which cause spectral broadening and this red shift [22].

R6G microcrystal spectrum (inset of figure 3) also shows band system with peaks at around 509 and 540 nm, red shifted with respect to the solution spectrum. Also the microcrystal spectrum shows a broadened spectral profile. This red shift along with the broadening is due to the presence of microcrystalline aggregate of R6G in the microcrystal films.

It is worthwhile to mention in this context that Arbeloa et. al. [23] reported the adsorption R6G molecules onto Laponite clay films at various concentrations and they showed that various H- and J-type dimer aggregates would form and coexist within the film. Grauer et al.[24] found red shift of about 10 nm for rhodamine 6G in Laponite surface. Tapia Estévez and co-workers [25, 26] reported a red shift of the monomer absorption of rhodamine 6G by 11 and 24 nm in Laponite.

In the present case the red shift of about 24 nm of R6G main absorption peak with respect to the corresponding solution spectrum and the close similarity to the microcrystal absorption spectrum is surely due to the formation of aggregation in the LbL films. These different aggregates may be possibly due to the electrostatic interaction between cationic parts of the dye molecules and the anionic parts of the surfactant SDS molecules i.e R6G molecules were subjected to more polar environments in the complex films [27, 28]. The ratio of the absorption intensity of these two bands (Figure not shown) is merely independent of concentration range of SDS solution used to fabricate PAH/SDS complex assemblies onto the solid substrates. In this case the electrostatic interaction dominate non-cooperative association between PAH and SDS where as hydrophobic moiety of SDS may favour aggregation by co-operative binding to the polymer. This also controls the overall aggregation of the dye molecules in the LbL films. This complexation may also depend upon the overall charge distribution of the polymer onto the quartz substrates.

The internal structure of the polyelectrolyte-Surfactant complex in self-assembled films is an important feature governing the further adsorption of the dye molecules to the terminal surface of complex film onto quartz substrates. In particular surfactant content and the degree of association between surfactant and PAH will determine the adsorption of R6G molecules as is evidenced from the sequential increase in the absorbance in the main absorption peak (540 nm) with increasing concentration of SDS in PAH/SDS complex molecular films. The effective adsorption requires available binding sites on the terminal layer. In such case the cationic part of the R6G molecules should interact with the available anionic binding sites of SDS in the PAH/SDS complex architectures.

To understand the kinetics of adsorption of R6G molecules onto the PAH/SDS complex assemblies, we have taken the absorption spectra of PAH/SDS/R6G complex films for different deposition times of R6G. Concentration of dye solution being $10^{-5}$ M of R6G. Here in all the cases the polycation (PAH) and anionic surfactant (SDS) deposition times were taken as 15 min but the dye deposition times were varied as 20, 30, 40, 50, 60, 300, 600, 900, 1200, 1500, 1800, 2400, 3000 and 3600 sec. After each immersion the substrate was washed off with an HCl aqueous solution and then dried by blowing nitrogen gas. It is observed from the figure 4(a) and 4(b) that the main absorption peak (540 nm) increases sharply up to 600 sec. and then remained merely constant for the deposition time greater than 600 sec. This is also evidenced from the plot of the the 540 nm band versus deposition time. It is important to mention that the interaction of R6G molecules to the PAH/SDS films was occurred so rapidly that almost 66% (as calculated from the change in absorbance intensity) of R6G molecules were deposited within first 20 seconds and then saturated after 600 sec. This observation reveals that the electrostatic adsorption of the dye molecules onto PAH/SDS complex film was completed after 600 sec. because of unavailability of the anionic binding sites in the complex architectures. Non-uniformity in the absorption maxima as evidenced from figure 4(b) occurs after 600 sec of deposition as the electrostatic interaction doesn't work as well and some repulsive type of interaction plays between the cationic part of the dye molecules in complex LbL films and the cationic group of the R6G dye molecules in the solution [29].

To visualize and investigate the formation of R6G domains as well as surface roughness in the PAH/SDS/R6G complex LbL self-assembled films we use Atomic Force Microscopy. Figure 5(a) and 5(b) show the AFM pictures of the PAH/SDS/R6G complex LbL film and PAH/SDS LbL films deposited onto glass substrate. Concentration of R6G aqueous solution was $10^{-5}$ M and its deposition time was 15 min. SDS Concentration was $10^{-4}$ M. From the figure 5(a) we observe closed packed conformation of dye molecules with their sharp edge and also their aggregates in the LbL films. This observation also confirms that the R6G molecules cover the whole area of the film after the interaction with SDS is completed. The RMS roughness of the film was calculated as 2.157 nm. Figure 5(b) shows the complex molecular assemblies of PAH/SDS onto glass substrate.

## 4. Conclusion

In summary, we have demonstrated the formation of Polymer/Surfactant complex molecular architectures onto the quartz substrates and the successful incorporation of cationic laser dye Rhodamine 6G onto the complex molecular system. These complexation and formation of dye aggregates are due to the electrostatic interaction between the oppositely charged species. Our observation reveals that in aqueous solution R6G molecules formed H-type aggregates (H-dimer) and these associates start to dominate on increasing the molar concentration as evidenced form their UV-Vis absorption spectroscopic data. However, in case of PAH/SDS/R6G complex LbL films the absorption bands have been shifted to 540 nm and 509 nm due to the formation of both H and J-type aggregates. Electrostatic interactions between the cationic dye and the anionic binding sites in the PAH/SDS facilitates complexation in the film. The ratio of the absorbance intensity of two bands in LbL film is merely independent of the concentration range of SDS solution used in the present work. Adsorption of R6G molecules to the PAH/SDS complex film was occurred so rapidly that 66% R6G deposited with 20 sec. and then saturated after 600 sec. of deposition as the unavailability of the anionic binding sites in the complex film after 600 sec. AFM picture also confirms the presence of R6G and their aggregates in the complex film.

**5. Acknowledgement**
Authors are grateful to Jadavpur University for providing financial assistance through JU Research grant (Ref. No. R-11/A/47/09) and excellent infrastructural facilities to perform this work. S.A. Hussain is grateful to DST for financial support to carry out this research work through DST Fast – Track project (Ref. No. SE/FTP/PS-54/2007).

**Figure Captions:**

Fig.1. Molecular structure of (a) Rhodamine 6G (R6G) (b) Poly(allyamine hydrochloride) (PAH) (c) Sodium dodecyl sulphate (SDS)

Fig. 2. Normalized UV-Vis absorption spectra of Rhodamine 6G in aqueous solution.

Fig.3. UV-Vis absorption spectra of PAH/SDS/R6G complex Layer-by-Layer self-assembled films for different varying concentrations of SDS aqueous solution ($10^{-2}$M-$10^{-5}$M of SDS). Concentration of R6G aqueous solution was $10^{-5}$M. Inset shows the R6G microcrystal absorption spectrum.

Fig. 4. (a) UV-Vis absorption spectra of PAH/SDS/R6G complex Layer-by-Layer self-assembled films for different deposition time of Rhodamine 6G onto PAH/SDS complex film. (Concentration of SDS and dye solution used $10^{-4}$M and $10^{-5}$M respectively) (b) Plot of absorbance intensity at 540 nm of PAH/SDS/R6G complex film verses deposition time.

Fig. 5: (a) Atomic Force Micrographs of PAH/SDS/R6G complex film deposited onto glass substrate. R6G deposition time was 15 min. (b) Atomic Fore Micrograph of PAH/SDS film deposited onto glass substrate.

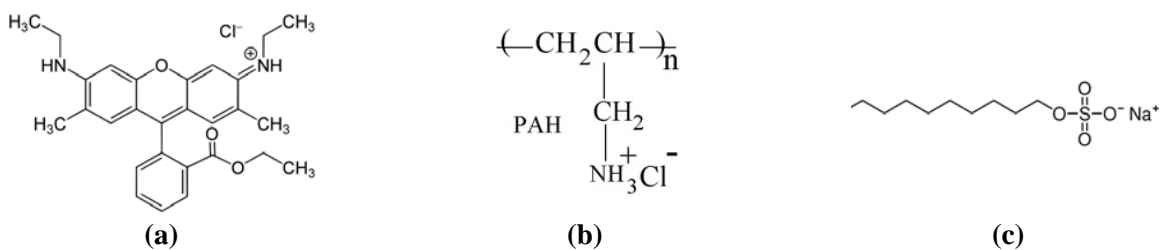

(a)     (b)     (c)

Fig. 1: P. K. Paul et.al.

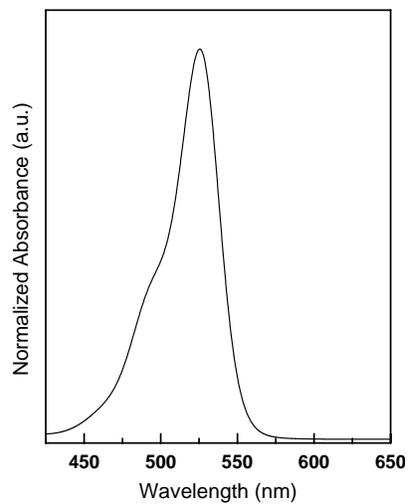

Fig. 2: P. K. Paul et.al.

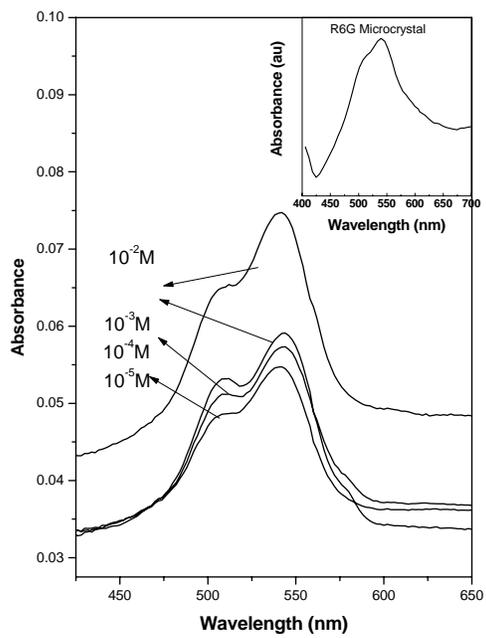

Fig. 3: P. K. Paul et.al.

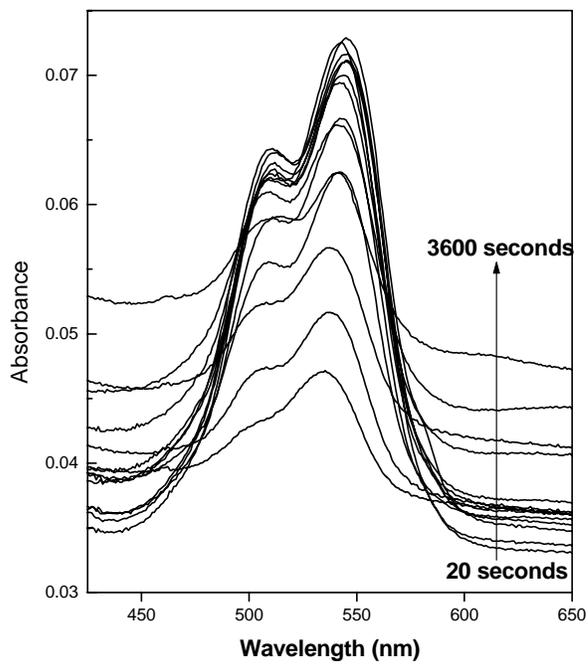
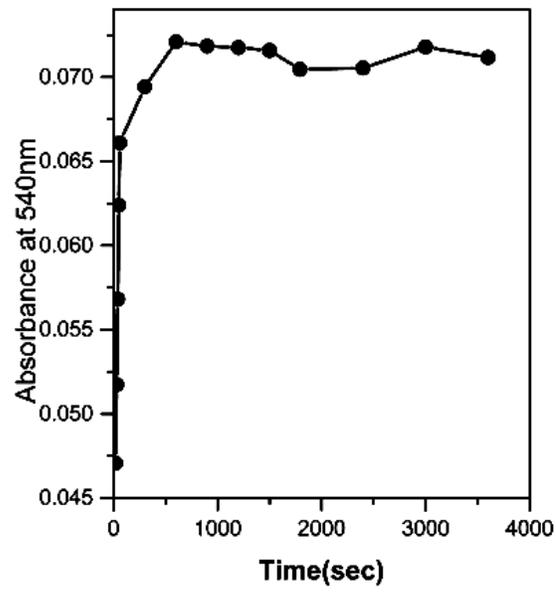

Fig. 4: P. K. Paul et.al.

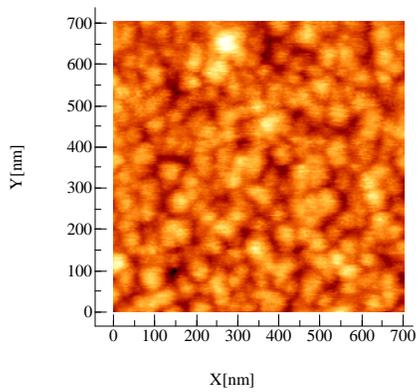
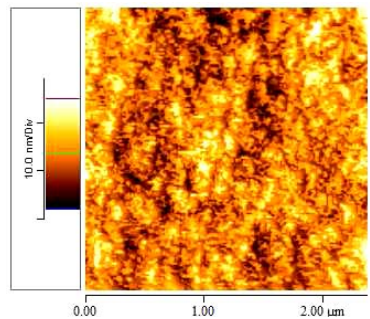

**(a)**                  **(b)**

Fig. 5: P. K. Paul et.al.

**8**